# STEREO observations of the energetic ions in tilted corotating interaction regions

R. Bučík [1], U. Mall [1], A. Korth [1], and G. M. Mason [2]

R. Bučík, Max-Planck-Institut für Sonnensystemforschung, Max-Planck-Str. 2, 37191 Katlenburg-Lindau, Germany. (bucik@mps.mpg.de)

U. Mall, Max-Planck-Institut für Sonnensystemforschung, Max-Planck-Str. 2, 37191 Katlenburg-Lindau, Germany.

A. Korth, Max-Planck-Institut für Sonnensystemforschung, Max-Planck-Str. 2, 37191 Katlenburg-Lindau, Germany.

G. M. Mason, Applied Physics Laboratory, Johns Hopkins University, Laurel, MD 20723, USA.

[1] Max-Planck-Institut für Sonnensystemforschung, Katlenburg-Lindau, Germany.

[2] Applied Physics Laboratory, Johns Hopkins University, Laurel, Maryland, USA.

**Abstract.**

In this paper we examine suprathermal He ions measured by the SIT (Suprathermal Ion Telescope) instrument associated with tilted corotating interaction regions (CIRs). We use observations of the two STEREO spacecraft (s/c) for the first 2.7 years of the mission, along with ground-based measurements of the solar magnetic field during the unusually long minimum of Solar Cycle 23. Due to the unique configuration of the STEREO s/c orbits we are able to investigate spatial variations in the intensity of the corotating ions on time scales of less than one solar rotation. The observations reveal that the occurrence of the strong CIR events was the most frequent at the beginning of the period. The inclination of the heliospheric current sheet relative to the heliographic equator (the tilt angle) was quite high in the first stage of the mission and gradually flattened with the time, followed by a decrease in the CIR activity. By examining the differences between measurements on the two STEREO s/c we discuss how the changes in the position of the s/c relative to the CIRs affect the energetic particle observations. We combine STEREO observations with observations from the ULEIS instrument on the ACE s/c and argue that the main factor which controls the differences in the ion intensities is the latitudinal separation between the two STEREO s/c relative to the tilted CIRs. The position of the s/c is less important when the tilt angle is high. In this case we found that the CIR ion intensity positively correlates with the tilt angle.

## 1. Introduction

Corotating interaction regions (CIRs) arise when fast solar wind, emerging from coronal holes, overtakes the slow solar wind emanating from the belt around the solar magnetic equator [*Gosling et al.,* 1993; *Balogh et al.,* 1999, and references therein]. The solar magnetic equator, tilted by an angle relative to the heliographic equator, maps outward into the heliosphere to form a surface, called the heliospheric current sheet (HCS). The heliospheric current sheet is a large-scale corotating structure in the heliosphere, which separates magnetic field lines of opposite polarities.

The interaction of the slow and fast solar wind flows leads to the formation of a pair of forward and reverse shocks at the boundaries of the CIRs, which occurs typically beyond the Earth's orbit [*e.g., Hundhausen and Gosling,* 1976]. These shocks can accelerate particles up to a few MeV/n in energy [*Fisk and Lee,* 1980]. In particular, energetic ions observed in the fast solar wind at 1 AU are expected to be energized at the CIR reverse shocks and propagate back toward the Sun along the magnetic field lines [*e.g., Barnes and Simpson,* 1976]. The observations of the suprathermal particles near 1 AU are inconsistent with the standard picture where the acceleration occurs by forward and reverse shocks. The peaks at low energies are observed within CIRs at places that are not magnetically connected to the shock [*Richardson and Zwickl,* 1984]. In addition, a turnover of the spectra below a few tens of keV/n predicted by Fisk and Lee's model is not observed [*Mason et al.,* 2008a]. These observational features suggest that ions are accelerated locally in unshocked compression regions [*Chotoo et al.,* 2000] either by a statistical mechanism [*e.g., Richardson,* 1985] or by compression acceleration [*Giacalone et al.,* 2002].

The observations made with instrumentation on board the Helios spacecraft (s/c) have shown that no recurrent fast solar wind streams and associated energetic particles were seen when the HCS is located very close to the heliographic equator [*Kunow et al.,* 1991]. The theoretical effort to explain such observations led to the development a three-dimensional CIR model for a tilted-dipole geometry at the Sun [*Pizzo* 1991; 1994]. *Pizzo* pointed out that during the declining and minimum phase of the solar cycle CIRs are organized around the HCS. The model suggests that CIRs are inclined with respect to the heliographic equator in the same sense as the HCS. At 1 AU the HCS is often found within a CIR. The model predicts that the strength of the compression region varies substantially with the latitude along the CIR. The highest compression is realized off the heliographic equator, tending to lie near the heliolatitudes roughly equal to the dipole tilt angle. In addition, the strength of the interaction regions changes with the tilt angle [*Gosling and Pizzo,* 1999]. If the tilt angle is high the stream-stream interaction is more direct and the compression regions are stronger. Conversely, when the tilt angle is small the interaction is weak and the CIR activity is confined to a narrow band around the heliographic equator.

Both latitude and tilt angle effects were seen in the solar wind plasma and energetic particle observations by the Ulysses s/c, which explored the three-dimensional heliosphere for the first time [*Simnett et al.,* 1994*; Gosling et al.,* 1997*; Sanderson et al.,* 1998, 1999*; Richardson et al.,* 1998; *Desai et al.,* 1999]. These surveys reported a latitude dependence of the reverse shock spectra with the strongest shock around the latitudes roughly equal to the tilt angle of the HCS. Furthermore, it was demonstrated that the particle intensity dependence on latitude may be expressed by an e-folding latitude of ~6°-8°, which also includes effects (minor) due to the change in the heliocentric distance between ~ 1 and 5 AU. *Sanderson et al.* [1999] examined the relation between the HCS and the recurrent particle increases associated with CIRs observed by the Ulysses s/c at 4-5 AU. These observations showed that the timing and amplitude of the peaks in the particle intensity profiles change as the dipole tilt changes, and as the warps in the current sheet change in amplitude. In addition, the authors concluded that the position of the s/c relative to the current sheet plays an important role in the observations of the CIR-related particles. Similar findings were earlier reported by *Sanderson et al.* [1998] with Wind observations in the ecliptic plane at 1 AU.

In this paper we investigate in detail the characteristics of energetic ions in tilted CIRs with the two STEREO spacecraft during the solar minimum of Cycle 23. Due to the generic coherence between CIRs and HCS we use the current sheet as a proxy to explore the latitude and tilt angle variations in the CIR related particles. We start with the discussion of the advantages of the configuration of the two STEREO orbits in Section 2. Section 3 presents an overview of the energetic ion observations. We compare the overall pattern of the helium ion increases with changes in the shape of the current sheet and further we identify CIR events for a detailed analysis. In Section 4 we examine differences between the occurrence of corotating ion increases on the two STEREO s/c. We combine the ion observations from STEREO and ACE to examine differences between ion intensities at three locations. We also investigate the relation between CIR ions and the tilt angle of the current sheet. Finally, in Sections 5 and 6 we discuss and summarize the results.

## 2. Experiment

Both STEREO s/c have heliocentric orbits near the ecliptic plane, STEREO-A preceding the Earth and STEREO-B trailing behind. This is achieved by placing the STEREO-A on an orbit closer to the Sun and STEREO-B further away from the Sun [*Kaiser et al.*, 2008]. Figure 1 (left side) shows the Heliocentric Earth Equatorial (HEEQ) positions of ACE, STEREO-A and STEREO-B s/c from January 2007 until October 2009. The separation of the two STEREO s/c increases in heliolongitude gradually in time (see panel A on right side of Figure 1). In October 2009 the heliolongitude difference was about 120° which corresponds to about 9 days of corotation time from STEREO-B to STEREO-A. The ACE s/c is in an orbit around the L1 point. The heliolatitude separation between the two STEREO s/c periodically changes with the amplitude increasing in time, as shown in panel (B) of Figure 1 (right side). In this panel the latitude difference between STEREO-B and STEREO-A is plotted by a black curve. Panel (B) also shows STEREO s/c heliolatitudes. The latitude separation between ACE and STEREO-A or STEREO-B changes in a similar manner with amplitudes around one half of the latitude separation between the two STEREO s/c. Another parameter which periodically changes with time is the radial distance from the Sun. Panel (C) of Figure 1 shows the heliocentric radial difference between STEREO-B and STEREO-A with a maximum of about 0.13 AU. Because of the unique configuration of their orbits, the two STEREO s/c can sample the same CIR twice at different heliolatitudes during periods which are shorter than one solar rotation. This can reduce the effects of the temporal variations in CIRs, for example, changes in the tilt angle of the HCS.

The measurements reported here were made with the Suprathermal Ion Telescope (SIT) [*Mason et al.*, 2008b] instruments which are part of the IMPACT suite [*Luhmann et al.*, 2008] on the STEREO-A and -B s/c. The SIT instruments on STEREO-A and -B have viewing directions close to the average

Parker magnetic field spiral line with a 44° field-of-view in the ecliptic plane. The SIT instrument is a time-of-flight mass spectrometer which measures ions from He to Fe in the energy range from 20 keV/n to several MeV/n. In our study we also make use of suprathermal He ion measurements made by the ACE/ULEIS instrument [*Mason et al.*, 1998], solar wind measurements made by the STEREO/PLASTIC instrument [*Galvin et al.*, 2008], and magnetic field measurements obtained by the magnetometer on STEREO [*Acuña et al.*, 2008]. In addition, we use ground-based observations of the solar magnetic field made by the Global Oscillation Network Group (GONG). We use data acquired over the extended minimum phase of Solar Cycle 23 from Carrington rotation (CR) 2052 to 2087 (January 2007-September 2009).

## 3. Observations and CIR Event Selection

Figure 2 provides an overview of the data during our investigated period. Panels (A) and (D) show suprathermal 1-hr average He ion intensities (particles/cm$^2$ s sr MeV/n) for 0.16-0.23 MeV/n measured by SIT-A and -B, respectively. The ion enhancements coincide with the high-speed streams shown in panels (B) and (E) indicating an association with CIRs. Vertical dashed lines in panels (A) and (D) mark the start times of the Interplanetary Coronal Mass Ejections (ICMEs) obtained from a list (STEREO Level 3 data) compiled by the STEREO magnetometer team at the University of California Los Angeles (http://www-ssc.igpp.ucla.edu/forms/stereo/stereo_level_3.html), based on plasma and magnetic field data. The list of the interplanetary shocks and CIRs observed by STEREO, provided by the same team, is also used in this paper. Black curves in panels (C) and (F) show the heliographic latitude of the current sheet seen from STEREO-A and -B, respectively, derived from the Sun's neutral line position provided by the GONG network (http://gong.nso.edu/data/magmap/). The heliographic latitude of the s/c is marked by a red line. Over-plotted by blue dots in the panels (C) and (F) is the magnetic field azimuth angle in Radial-Tangential-Normal (RTN) coordinates with the scale on the right y-axis. The vertical red lines in panels (C) and (F) mark the start times of the CRs from 2052 to 2088.

The solar neutral line provided by GONG is calculated from the photospheric field magnetograms from a six-station network using the Potential-Field Source-Surface model of the coronal magnetic field [*Altschuler and Newkirk,* 1969]. For the particular Carrington rotation the GONG data contain the heliolatitude of the neutral line with 1° step in heliolongitude. We converted this longitude by linear interpolation to the time, where 360° corresponds to the start time of the CR. Start times of the Carrington rotations for STEREO-A and -B were determined from the start times of the CRs for Earth, considering the angular offset of the STEREO s/c from the Earth-Sun line. The results were delayed by five days, to represent the current sheet at 1 AU. This delay corresponds to a propagation speed of ~350 km/s. Earlier investigations reported that the simple constant-speed mapping technique is a good approximation of the current sheet at 1 AU [*Sanderson et al.*, 1998; *Lee et al.*, 2010]. Careful inspection of the panels (C) and (F) in Figure 2 reveals that the crossings of the inferred current sheet (intersections of the red and black curves) closely match the reversals in magnetic field polarity observed by STEREO in-situ measurements over the whole period of the present study.

Further exploration of Figure 2 shows that during the whole investigated period the two STEREO s/c regularly cross or approach the current sheet at different heliolatitudes. We divide this period into the five different intervals, based on the current sheet shape. In the interval (I) the HCS was warped. A warp comparable to the main current sheet latitude extent was visible at the beginning of the period with the amplitude slowly decreasing in time. We observe that the rate of occurrence of the ion enhancements, as well as the ion intensities were relatively high. In the period (II) the HCS was almost flat with a somewhat higher heliolatitude compared to the previous period. The rate of occurrence of the CIR events and interestingly peak intensities as well subsided somewhat. In the interval (III) the distance of the HCS from the heliographic equator notably decreased and small distortion in the current sheet developed again. The peak intensities as measured by STEREO-A are somewhat decreased, but only a small change is seen on STEREO-B. In period (IV) the HCS was again flat with a heliolatitude temporally increased. This period was quiet in the CIR event activity, notably on STEREO-A. In period (V), the current sheet was almost aligned with the heliographic equator, but no marked difference in particle intensities compared to period (IV) was observed. The occurrence rate and intensities of the ion enhancements are very similar on the two s/c in interval (I), when STEREO-A and STEREO-B trace almost the same latitudes relative to the current sheet. In interval (II) the phase shift between s/c latitudes starts to increase. It is accompanied by the differences in the pattern of particle increases between STEREO-A and –B, wherein CIRs observed on one spacecraft are often not seen on the other.

In our study we use the list of the CIR events observed by STEREO and ACE in the interval January 2007 - September 2008 which has been compiled in a study of *Mason et al.* [2009]. The authors required that the hourly averaged helium ion intensity in the 0.16-0.23 MeV/n energy range exceed 10 particles/(cm$^2$ s sr MeV/n) on at least one of the two STEREO s/c (see dashed horizontal lines in Figure 2A and 2D). They used this criterion to obtain reasonable statistical accuracy for studying the spectral characteristic in the CIR ions. We use the same selection criterion for CIR enhancements through October 2008 and September 2009.

Table 1 lists pairs of the corresponding CIR ion events observed on STEREO-A and STEREO-B. The leftmost column is the CIR number. The events 1 – 35 were identified by *Mason et al.* [2009]. Columns 2 and 6 list the approximate start times of the events either from the survey of *Mason et al.* [2009] (events 1-35) or identified for the remaining events in this study. In case of low ion intensities, where the start of the event is unclear, we denote the event by the start of the solar wind speed rise. Columns 3 and 7 list the 1-hr event maximum He intensity in the energy range 0.16-0.23 MeV/n measured by STEREO-A and STEREO-B, respectively. The time at the peak intensity is shown in Columns 4 and 8. Columns 5 and 9 indicate the s/c heliolatitude at the time of the peak intensity. Column 10 shows the Carrington rotation number. The corotation delay time between spacecraft, calculated from the s/c heliolongitude difference divided by the Sun's rotation rate, is shown in Column 11. Column 12 has some notes on the CIR events. The events marked by the letter (x) have 1-hr intensity < 1 particle/(cm$^2$ s sr MeV/n) which corresponds to about 6 counts in 1 hour. An instrumental background (~ 2 counts/1-hr) in the energy range 0.16-0.23 MeV/n on STEREO-A was subtracted from the measured ion intensities. The letters (a) and (b) indicate events with local reverse shocks. Only shocks with signatures in both magnetic and plasma solar wind data are listed in the table. The table is divided by horizontal lines into the five parts corresponding to the periods (I) to (V). Table 2 lists the CIR ion events observed on ACE. The columns show the CIR number, the approximate start time of the event, the 1-hr event maximum He intensity in the energy range 0.16-0.23 MeV/n measured by ULEIS, the time at the peak intensity, and the s/c heliolatitude at the time of the peak intensity, respectively.

In this survey we identify pairs of the corresponding CIR events on the basis of the nominal corotation delay between STEREO-A and -B by inspection of the ion intensity time profiles. This method has been used by *Richardson et al.* [1998], for association of the Ulysses corotating events with the particle events at the Earth observed by IMP 8. For the cases when the ion intensities were very low we inspect the solar wind speed time profile and use the list of the CIRs in the STEREO Level 3 data.

Figure 3 illustrates the identification of the CIR events. The figure shows two pairs of the CIR events (#40 and #41) with very low ion intensities on STEREO-A. Upper four panels (A, B, C, and D) show for STEREO-A from top to bottom 1-hr He ion intensities in four energy channels, solar wind speed, total pressure (sum of magnetic and plasma proton pressures), and magnetic field azimuth in RTN coordinates. The lower panels (E, F, G, and H) show the same parameters for STEREO-B. Middle values of the energy ranges for He ions in the units of MeV/n are shown right to the second y-axis. The pairs of the vertical red lines indicate the CIR boundaries found in the list of the CIRs in the previously cited UCLA STEREO level 3 data. The blue slant lines indicate the corresponding corotating increases on STEREO-B and STEREO-A. The time shift between the maximum intensities on STEREO-A and STEREO-B is for both CIRs about 6-6.5 days. This value is not far from the nominal corotation delay of 6.9 days (see Table 1). The approximate times of the HCS crossings are shown by black lines in panels (D) and (H). We can see that the corresponding intensity enhancements occur in the same magnetic sectors. This suggests that these enhancements would be associated with the same CIR, though *Reames et al.* [1997] reported that at the higher energies (> 1 MeV/n) a very late phase of the CIR event may continue in the opposite magnetic sector. We note that the majority of the CIR events in period (I) were associated with Southern Hemisphere CIRs. In the interval (II) the numbers of southern and northern CIR events were approximately the same. The CIR events in periods (III) to (V) were mainly due to Northern Hemisphere CIRs.

## 4. Results

In the following we investigate in more detail the latitude and tilt angle effects on energetic particle increases by examining the differences between observations on STEREO-A and STEREO-B. To avoid any possible interference between the CIRs and the transient disturbances driven by ICMEs [*Crooker et al.*, 1999], which might intensify the strength of the compression regions [*Keppler et al.*, 1995] the pairs of particle events with mixed ICME-CIR at least on one s/c (14 out of 47 pairs) are not considered in the following analysis. For two pairs of the events (events #9 and #19) the ICME was present on both STEREO-A and -B. Thus there are 16 mixed ICME-CIR ion events. From this number, in 8 cases the ICMEs were embedded in or overlapped with CIRs, 5 times were ICMEs < 0.5 day away from CIRs, and in 3 cases the ICMEs were ~ 1-2 days away from CIR. We note from Figure 2 that some of the most prominent He peaks occur around the times of the ICMEs. For example, 4 (3) out of 8 CIR events with intensities greater than 100 particles/($cm^2$ s sr MeV/n) were accompanied by ICME on STEREO-A (-B). It is not coincidental that the STEREO s/c see the ICMEs around the CIR events. An intimate association between ICMEs and CIRs derives from the fact that CMEs are commonly emitted from the belt of the slow solar wind [*Crooker et al.*, 1999]. We see that in some cases ion intensity peaks in mixed events are not very strong, which may be related to different properties (e.g., propagation speed) of ICMEs.

### 4.1. Rate of the Occurrence of CIR Enhancements

Figure 4 shows the helium ion SIT instrument data in a different form than Figure 2. In Panels (A) and (C) of Figure 4 the color scale marks a logarithm of 0.16-0.23 MeV/n energy helium intensities versus Carrington longitude between the CR 2053 and CR 2087 rotations. In Carrington coordinate system, longitudes are fixed on the surface of the Sun. Panel (A) shows data for STEREO-A and panel (C) data for STEREO-B. The counts are collected in Carrington longitude bins with the width of 1°, which roughly correspond to 1.8 hours. Our data show three recurrent source regions (strips of enhanced number of counts) responsible for the CIR events in period (I), two in period (II) and one stable recurrent region in period (III). In periods (IV) and (V) we observe an irregular pattern in the distribution of the number of counts. Furthermore, only little difference is seen in the structure between data on the two s/c in period (I) and in the most part of period (II). From the end of period (II) the difference start to be more striking. For example, the enhancement in CR 2069 around 160° of Carrington longitude is seen only on STEREO-B; or the enhancement in CR 2070 observed by STEREO-B between 80°-110° is missing on STEREO-A. Finally, we note that the strips of enhanced number of counts are inclined toward the east relative to the Sun's rotation. This indicates that energetic ion increases recur with periods which are slightly shorter than one Carrington rotation. The drift of CIR source regions in Figure 4 is due to the fact that the Carrington rotation is set for mid latitudes, while we are viewing in the ecliptic and so are sampling a faster rate.

Panels (B) and (D) in Figure 4 show histograms of the number of CIR enhancements in Carrington rotations observed by STEREO-A and STEREO-B, respectively. Only He ion enhancements with an intensity > 10 particles/($cm^2$ s sr MeV/n) in the range of 0.16-0.23 MeV/n on each s/c are taken into account. Panel (E) shows the HCS tilt angle (black line). The tilt angle is the same as the maximum heliolatitude attained by the HSC. As defined by *Hoeksema* [1991] we calculate the tilt angle as an average of the maximum northern and southern latitude extent of the HCS during each CR. The latitudinal extension of the CIR is similar to the latitudinal extension of the HCS, whereas the reverse shocks may extend beyond the latitude tips of the HCS at greater heliocentric distances [*Gosling et al.*, 1993]. In this preliminary survey the HCS tilt angle may serve as an indirect estimate of the inclination of the CIRs and gives us a clue on the latitude extension of the compression region. We note that *Riley et al.* [1996] analyzed the solar wind flow pattern to infer the local orientation of the CIRs.

Panels (B) and (D) in Figure 4 show two CIR events per CR only during the period of enhanced tilt angles. In intervals from (IV) to (V) there exist periods lasting for eight (STEREO-A) or four (STEREO-B) solar rotations without any CIR activity with ion intensities above 10 particles/($cm^2$ s sr MeV/n). The relative number of events, shown by solid (STEREO-A) and dotted (STEREO-B) blue lines in panel (E), decreases from interval (I) to interval (V). Through the period (II) to (V) this decrease follows the tilt angle. The enhanced rate of the occurrence of the CIR ion intensity increases could also be related to large warps in the current sheet, which may give rise to other interaction regions. From Figure 2 it may be seen that some warp crossings in the interval (I) are associated with strong helium peaks.

A comparison of the histograms in panels (B) and (D) in Figure 4 shows there are periods with duration of five CRs in interval (I) with no differences between the number of events on STEREO-A and -B. In contrast to the interval (I), the period (IV) shows differences for every CR. Red crosses in panel (D) mark the CRs with a different number of corotating increases on the two STEREO s/c. There are 25 %, 30 %, 50 %, 100 % and 20 % of such CRs in the intervals (I), (II), (III), (IV) and (V), respectively. These observations indicate the large differences between the number of events on STEREO-A and STEREO-B in a period of low tilt angles and little difference when the tilt angle increases to higher values. An exception is interval (V), where due to weak intensity increases the differences in the number of events above the threshold intensity could not be properly determined.

### 4.2. Relation to the Tilt Angle

Figure 5 shows 1-hr He ion CIR event peak intensities in the 0.16-0.23 MeV/n energy range observed by SIT-A (open symbols) and SIT-B (solid symbols) as a function of the HCS tilt angle during the rotation when the CIR occurred. The data points labeled by S refer to the CIRs bounded by in-situ reverse shocks. The two arrows point to the CIR event on 29 January 2007 (event #1) which was preceded by a $^3$He-rich solar energetic particle (SEP) event on 24 January 2007. The 24 January 2007 event was the largest $^3$He-rich event detected from 2007 through mid-2009 [*Wiedenbeck et al.*, 2010].

In period (I)-(crosses) when the current sheet was warped and the tilt angle high, a very broad range of ion intensities is observed in a narrow range of tilt angles, which is highlighted by the shaded area in Figure 5. The events with the high intensities (> 100 particles/cm$^2$ s sr MeV/n) in the interval (I) are either events with in-situ-shocks or events preceded by a $^3$He-rich SEP event. A positive correlation between tilt angle and ion peak intensity is observed in period (II)-(triangles) when the current sheet is flat and tilted to the high angles (~ 32°). The linear correlation coefficient ($r$ = 0.68) is statistically significant with a low probability ($p$ ~ 0.4 %) resulting from a random population. The probability $p$ was calculated using a Student's distribution function [*Cooper*, 1969]. The solid line in the figure presents a least-square fit of ln(intensity) versus tilt angle for the CIR events from interval (II). In periods (III)-(circles), (IV)-(squares), and (V)-(bowtie) when the tilt decreases to values < 20° no clear relation between the ion intensity and the tilt angle is seen.

### 4.3. Intensity Differences on STEREO-A and -B

In this section we examine the differences between the ion intensities on the two STEREO s/c with respect to the changes in latitude separation between the two s/c. Due to latitude gradients the two s/c which are located at different heliographic latitudes cross the CIRs at different strengths of the compression region. From simple geometric considerations it follows that as the tilt angle decreases the s/c approach to the north-south edges of the CIR. At low tilt angles, the two s/c even with small latitude separation may cross regions with very different compressions. To account for such an effect, we use the latitude differences between the two s/c divided by the tilt angle.

Figure 6 shows the time profile of the ratios between 1-hr CIR event peak intensities on STEREO-A and STEREO-B for He ions in the 0.16-0.23 MeV/n energy range. The ratios are computed in the way that the higher intensity is divided by the lower one. The CIR events for intervals (I), (II), (III), (IV) and (V), as determined in Figure 2, are shown by different symbols. The open symbols indicate cases where a CIR was bounded by the in-situ reverse shock at least on one s/c. Corresponding s/c latitude differences relative to the tilt angle ($\Delta Lat/Tilt$) are marked by the X-symbols. The absolute latitude separation between the two STEREO s/c for the events in Figure 6 is between 0.02$^{o}$ and 11.4 $^{o}$, which corresponds to distances between $10^{-4}$ AU and $10^{-1}$ AU. The values of the parameter $\Delta Lat/Tilt$ show large differences in scale. We used a logarithmic scale for the variable $\Delta Lat/Tilt$ since 48% of the values are between 0.01-0.1 and 39% between 0.1-1.0. All quantities in Figure 6 are plotted at times of peak intensities on STEREO-A.

There are several interesting aspects of the data shown in Figure 6. One interesting feature seen in Figure 6 is that the intensity ratio time profile is very similar to the time profile of the latitude difference in the period (I), (II) and (III). Through these three intervals the ratio periodically increases and decreases following the changes in the time profile of the latitude differences. In periods (IV) and (V) the time profile of the intensity ratios is unclear due to high statistical errors. For the data point with open symbol in interval (III), event #34, the peak intensity factor of seven higher was associated with a CIR accompanied by an in-situ shock, compared to the same CIR with no in-situ shock. For other events indicated by open symbols the ion intensity in CIR events accompanied by in-situ shocks was either lower or comparable to the case without in-situ shocks.

We note from Figure 6 that in the interval (II), in the period between September 2007 and February 2008, the intensity ratio does not change a lot and remains in the range of ~ 1-2. Figure 1C shows that in the same time period the radial separation between the two s/c decreases. In the interval (I) when the intensity ratios show increase (period March - July 2007) the radial separation between the two s/c increases as well. This immediately brings up the question whether the variations in the intensity ratios are related to the changes in the s/c radial differences. It has been reported earlier [*van Hollebeke et al.*, 1978] that CIR particle intensity increases with radial heliocentric distance. In heliocentric radial range 0.46-2.2 AU the positive radial intensity gradient of 300%/AU has been reported. Thus it may be expected that the difference between peak intensities on two s/c would increase (decrease) with the increasing (decreasing) radial distance between the s/c.

The correlations between the intensity ratios and latitude differences relative to the tilt angle and between the intensity ratios and radial differences are examined in Figure 7. The data in Panel (A) of Figure 7 show a positive correlation for s/c latitude differences relative to the tilt angle. The value of the linear correlation coefficient $r$ = 0.60 along with its statistical significance ($p$ ~ 0.02 %) indicate that the differences between He ion intensities observed on STEREO-A and STEREO-B are well ordered with the latitude difference between the two s/c relative to the tilt angle. As the s/c latitude separation relative to the value of the tilt angle increases, the differences between peak intensities on the two STEREO s/c become larger. This trend is shown by the solid line, which represents a linear least-square fit to the data. The correlation coefficient (*r* = 0.61) for the CIR events in time period from (I) to (III) remains high, indicating a low probability (*p* ~ 0.06 %) of resulting from random population. Panel (B) in Figure 7 shows no positive correlation between the ion intensity ratio and the s/c radial difference. The correlation coefficient (*r* = -0.04) has a high probability (*p* ~ 82 %) that these two populations are unrelated.

**4.4. Intensity Differences between STEREO and ACE**

A comparison between the He ion intensity differences on the two STEREO s/c and the intensity differences on ACE and STEREO is shown in Figure 8A. The SIT instruments were cross-calibrated with ULEIS during the first part of 2007 when all three s/c were close to each other [*Mason et al.,* 2009]. Figure 8A shows 1-hr He peak intensity ratios for SIT-A and SIT-B versus the 1-hr peak intensity ratios for SIT-A and ULEIS (blue symbols) and SIT-B and ULEIS (red symbols) in the energy range 0.16-0.23 MeV/n. The points are mostly spread above the diagonal line indicating that for the same CIR event the differences between peak intensities on the two STEREO are higher compared to the differences between peak intensities on ACE and STEREO. Four outlier points below the diagonal line correspond to the CIR events #21 (triangles) and #31(circles). In these events He peak intensities observed by ACE were about one order of magnitude higher than intensities seen by both STEREO-A and STEREO-B. The enhanced intensity in event #31 could be explained by particle acceleration in the local reverse shock detected by ACE, but not by STEREO. No local reverse shocks were seen in event #21 on any of the three spacecraft (L. Jian, private communication, 2011). Thus the high He intensity measured by ULEIS/ACE in event #21 is due to other causes.

Figure 8B plots *ΔLat/Tilt* for the two STEREO s/c versus *ΔLat/Tilt* calculated for STEREO-A and ACE (blue symbols), and STEREO-B and ACE (red symbols). It is seen that in almost all CIR events the latitude separation between the two STEREO s/c is larger than the latitude separation between ACE and STEREO. For 3 out of 4 CIR events in interval (IV) the latitude separation between ACE and STEREO-A, or ACE and STEREO-B was equal or higher compared to the latitude separations between the two STEREO s/c (see squares below the diagonal line). Figure 8A shows that for all events in interval (IV) the difference between the intensities on the two STEREO s/c are larger than between ACE and STEREO. The data points marked by circles and crosses below the diagonal line correspond to events #33 and #4. For these events the intensity difference on ACE and STEREO was higher than the intensity difference on the two STEREO s/c.

## 5. Discussion

During the January 2007 – September 2009 solar minimum period STEREO observations revealed that the occurrence of the strong CIR events was the most frequent in the period when the tilt angle of the HCS was high and when large warps in the current sheet were developed. A gradual decreasing of the tilt angle with the time was followed by decrease in the CIR activity. These findings are consistent with the results reported by *Sanderson et al.* [1998], who found that the tilt and the distortions in the current sheet play a major role in explaining the frequency of the occurrence of the recurrent compression regions. The high occurrence rate of the CIR events during periods of high tilt angles may be interpreted as a result of the dependence of the strength of the compression and consequently, the particle intensity on the tilt angle. Thus, where the tilt angle is high the events are strong enough to exceed the He ion intensity threshold selected in this analysis. Our observations show that large differences in the occurrence rate of CIR events are seen during periods of low tilt of the current sheet when the two STEREO s/c were largely separated both in heliographic longitude and latitude.

As previously discussed in the literature [*e.g., Kocharov et al.,* 2003], the energetic particles from SEP events could be further accelerated by CIRs. Often the intensity of the recurrent energetic particles was enhanced after the solar transient events [*e.g., Sanderson et al.,* 1995]. *Kocharov et al.* [2008] have analyzed the relationship between $^3$He-rich SEP events and solar wind high-speed streams. The authors suggest that energetic ions from impulsive solar events can be temporary confined in the corotating compression regions in solar wind. This may explain the observations of the enhanced ion intensities in the CIR event on 29 January 2007 (event #1) which was preceded by the $^3$He-rich SEP event. A strong CIR event on 24 May 2007 (event #9), preceded by two small SEP events on 19 and 23 May 2007 [*Bučík et al.,* 2009] is not considered in this analysis primarily due to transit of the ICME. Thus there were two other SEP events in May 2007 which may also fill up the heliosphere, providing the seed particles for CIRs to be accelerated in the earlier parts of the investigated period and causing enhanced ion intensities in the interval (I).

We note that during the stages of the mission when the tilt angle was relatively small, < 20°, the STEREO observations do not show a relation between tilt and the particle intensity, while above roughly 30° the peak intensities show quite good organization by the inferred tilt angle of the current sheet. These observations are consistent with the results obtained from plasma and magnetic field data from Ulysses. *Gosling et al.* [1993] reported that the predictions of the tilted-dipole model of the CIRs are in excellent agreement with the observations at heliocentric distance of ~5 AU during the modest tilt of ~30°. *Gosling et al.* [1995] noted that less conclusive results were obtained during the traverse across the equator at a distance of 1.3 AU when the tilt angle was ~10°. They suggested that some of the complexity in the flow pattern associated with CIRs they observed was either the consequence of the small tilt or the warping of the current sheet. Furthermore, simulations of the *Riley et al.* [2002] have shown that near solar minimum, the HCS is better described as a flat surface with warps in it, in contrast to the sinusoidal picture that is generated by considering the interplanetary extension of the tilted dipole. The complex pattern of the corotating particle intensities near solar minimum has been reported by *Desai et al.* [1998] from Ulysses observations. The authors suggest that the irregularities in the CIR events were caused by the warps in the HCS which was in that time almost aligned with the heliographic equator.

In the periods of low tilt angles, intervals (III)-(V), the STEREO s/c heliolatitude becomes closer to the maximum latitude extension of the heliographic current sheet, where the compression regions should be the strongest. Such conditions could lead to the enhanced particle intensities observed in the later part of the investigated period. The lack of the organization by tilt angle is presumably related to the latitude variations in the CIRs which STEREO could easier experience when the tilt angle is small. In contrast to intervals (III) to (V), the s/c latitudes in interval (II), present only a fraction of the maximal latitude extension of the CIRs. Therefore, the latitude effects are small in interval (II). In spite of enhanced values of the tilt angle in interval (II), being out of the region of the maximal compression may cause that the measured intensities in some events are relatively low. *Leske et al.* [2009] have also discussed the relation between STEREO observations and the HCS.

For most of the investigated periods, (intervals (I) - (III)), the differences between He ion peak intensities on STEREO-A and -B are systematically ordered by the s/c latitude separation relative to the tilt angle. In the intervals (IV) and (V), when the spacecraft were well separated in the longitude, the temporal variations in the strength of the CIRs can not be ruled out. *Mason et al.* [2009] discussed the effects of the s/c latitude connection to the source coronal holes on differences in energetic particle observations between the two STEREO s/c. The authors noted that since many of the events were recurring the bulk of the changes were not due to temporal evolution of the CIRs. We remind that CIR events in the survey of *Mason et al.* [2009] are from the periods (I) - (III) identified in this study (see Table 1). Indeed, the color diagrams in Figure 4 (panels A and C) show two strips of enhanced counts, one that persisted for ~ 25 solar rotations in the intervals (I) - (III) and other for ~ 16 rotations in the intervals (I) - (II). Lacking of any stable pattern in the intervals (IV) and (V) suggests that the temporal changes in the CIRs may play a role in the later stage of the investigated period. However, to which extent the temporal variations contribute to the ion intensity differences in the intervals (IV) and (V) on the time scales about one quarter of the solar rotations is not clear from these investigations.

The combination of STEREO and ACE observations shows that, in general, the intensity differences between the two STEREO s/c were larger than the intensity differences between ACE and STEREO. The latitude separations between the two STEREO s/c were larger than between ACE and STEREO with exception of some events in the later stage of the investigated period. This finding is consistent with the suggestion that the position of the s/c relative to the latitudinal extent of the CIR is the key factor determining the observed differences between the CIR-related increases at the different locations. The observations show that for the CIR events in the interval (IV) the intensity differences between the two STEREO s/c were higher compared to the intensity differences between ACE and STEREO although the latitude separations were lower. This suggests that near the end of the investigated period, when the two STEREO s/c were widely separated in longitude, the differences between CIR enhancements on the two STEREO s/c have to be explained by factors different from latitude separation. The temporal changes in the source regions for the CIRs during the corotation time between the two STEREO s/c would be the most plausible explanation.

For the range of radial differences between the two STEREO s/c the intensity of the He ions in the CIR events on STEREO-B should be on average 1.3 times higher compared to the intensities on STEREO-A, assuming positive radial gradient of ~ 300%/AU. The factor of 1.3 is probably too low for radial separation to play an important role in the intensity differences between the two STEREO s/c, which is manifested in a lack of the correlation between these two variables.

The fact that in some events suprathermal ions are accelerated in shocks developed at ~ 1 AU will introduce another variable. Due to propagation effects, such as adiabatic deceleration in the expanding solar wind [*Fisk and Lee*, 1980], the flux of the particles accelerated in shocks beyond the Earth’s orbit should decrease in a given energy range while streaming back to 1 AU.

## 6. Conclusions

We have examined suprathermal He ions in the 0.16-0.23 MeV/n energy range from STEREO during the minimum of Solar Cycle 23. The main results can be summarized as follows:

1. The high occurrence rate of the CIR ion enhancements is observed during periods of high tilt angles and large warp in the current sheet. The rate decreased with the current sheet tilt angle. The differences between the number of CIR events on STEREO-A and STEREO-B were very small when the tilt angle was high. Conversely, during periods of lower tilt angles these differences increased.

2. The combination of the observations on the two STEREO s/c with the observations on the ACE s/c demonstrates that the CIR intensity differences between the spacecraft are well ordered with the corresponding latitude differences relative to the tilt angle. Temporal changes in the CIRs might have contributed to the intensity differences between the two STEREO s/c in the end of the investigated period which is characterized by the absence of the stably recurring corotation source regions for the CIRs.

3. The He ion CIR intensity is positively correlated with the tilt angle during periods of a highly inclined current sheet. No clear relation between ion intensity and tilt angle is observed in the period of low tilt angles. The lack of the correlation is likely related to the effects of latitude intensity variations in the CIRs.

**Acknowledgements.** We acknowledge the STEREO PLASTIC team (NASA Contract NAS5-00132) for the use of the solar wind plasma data. This work utilizes data obtained by the Global Oscillation Network Group (GONG) program, managed by the National Solar Observatory, which is operated by AURA, Inc. under a cooperative agreement with the National Science Foundation. The data were acquired by instruments operated by the Big Bear Solar Observatory, High Altitude Observatory, Learmonth Solar Observatory, Udaipur Solar Observatory, Instituto de Astrofísica de Canarias, and Cerro Tololo Interamerican Observatory. This work was supported by the Max-Planck-Gesellschaft zur Förderung der Wissenschaften and the Bundesministerium für Wirtschaft under grant 50 OC 0904. The work at the Johns Hopkins University/Applied Physics Laboratory was supported by NASA grant NNX10AT75G for ACE and contract SA4889-26309 from the University of California, Berkeley for STEREO. We thank T. Wiegelmann for helpful discussions on the large scale magnetic field of the Sun. We would also like to thank B. Podlipnik for his support in the computer and network areas.

**Figure 1.** Left side: Heliographic Earth Equatorial (HEEQ) positions of STEREO-A (red trace) and STEREO-B (blue trace) from 1 January 2007 through 1 October 2009. The ACE s/c position corresponds to the black cross mark connecting the red and blue curves. The Archimedes spiral lines are for 350 km/s solar wind speed. Right side: Panel (A): Heliographic longitude difference (left y-axis) and delay time (right y-axis) between STEREO-B and STEREO-A. Panel (B): Heliographic latitude difference between STEREO-B and STEREO-A (black curve). STEREO-A (red curve) and -B (blue curve) heliolatitudes. Panel (C): Heliographic radial difference between STEREO-B and STEREO-A.

**Figure 2.** Panel (A): SIT/STEREO-A hourly averaged suprathermal He intensities (particles/cm$^2$ s sr MeV/n) in the energy range 0.16-0.23 MeV/n. Panel (B): Solar wind speed from STEREO-A. Panel (C): Heliographic latitude of the current sheet for STEREO-A (black trace), heliographic latitude of STEREO-A (red trace), and magnetic field azimuth in RTN from STEREO-A (blue dots). Panel (D): SIT/STEREO-B hourly averaged He intensities (particles/cm$^2$ s sr MeV/n) in the energy range 0.16-0.23 MeV/n. Panel (E): Solar wind speed from STEREO-B. Panel (F): Heliographic latitude of the current sheet for STEREO-B (black trace), heliographic latitude of STEREO-B (red trace), and magnetic field azimuth in RTN from STEREO-B (blue dots). Roman numbers in panel (A) refer to the different periods (separated by solid blue vertical lines) identified in the text. Dashed vertical black lines in panels (A) and (D) indicate start times of the ICMEs. Red vertical lines in panels (C) and (F) mark start times of the CRs from CR 2052 to CR 2088. Red triangles mark from left to right the CR 2060, 2070 and 2080.

**Figure 3.** Panel (A): SIT/STEREO-A hourly averaged suprathermal He intensities (particles/cm$^2$ s sr MeV/n) in the energy channels 0.189, 0.268, 0.384, and 0.550 MeV/n. Panel (B): Solar wind speed

from STEREO-A. Panel (C): Total pressure from STEREO-A. Panel (D): Magnetic field azimuth angle in RTN coordinates from STEREO-A. Panels (E, F, G, and H) are same as panels (A, B, C, and D) but for STEREO-B. Black vertical lines in panels (D) and (H) mark approximate time of the HCS crossings deduced from the abrupt change in the magnetic field azimuth angle. Red vertical lines mark boundaries of the CIRs. Blue slant lines mark the two pairs of the CIR events on STEREO-A and STEREO-B.

**Figure 4.** Panel (A): SIT/STEREO-A He counts in the energy range 0.16-0.23 MeV/n in Carrington longitude *vs.* Carrington rotation number diagram. Panel (B): Histogram of the number of He ion enhancements in Carrington rotations for STEREO-A. Panel (C): SIT/STEREO-B He counts in the energy range 0.16-0.23 MeV/n in Carrington longitude *vs.* Carrington rotation number diagram. Panel (D): Histogram of the number of He ion enhancements in Carrington rotations for STEREO-B. Red crosses are described in the text. Panel (E): HCS tilt angle. Numbers indicate medians of the tilt in intervals (I)-(V). Number of events per Carrington rotation in intervals from (I) to (V) for STEREO-A (solid blue lines) and STEREO-B (dashed blue lines). In interval (II) and (III) two blue lines overlap each other.

**Figure 5.** 1-hr CIR event peak intensity for 0.16-0.23 MeV/n He ions at STEREO-A (open symbol) and STEREO-B (solid symbols) *vs.* HCS tilt angle. The CIRs with in-situ reverse shocks are labeled by S. Two arrows mark events preceded by $^{3}$He-rich SEP events. The quantities $r$ and $p$ denote the linear correlation coefficient and its statistical significance for the data in interval (II). Solid line is a linear least-squares fit to the data in interval (II).

**Figure 6.** Ratio between 1-hr CIR event peak intensities on STEREO-A and STEREO-B for 0.16-0.23

MeV/n He ions (left scale). The symbols refer to the different periods identified in the text. Open symbols indicate events with the in-situ reverse shocks. Over plotted by X-symbols are heliographic absolute latitude differences between the s/c divided by the tilt angle (right scale).

**Figure 7.** Panel (A): Ratio between 1-hr CIR event peak intensities on STEREO-A and STEREO-B for 0.16-0.23 MeV/n He ions *vs.* heliographic latitude difference between the s/c relative to the tilt angle. Panel (B): Ratio between 1-hr CIR event peak intensities on STEREO-A and STEREO-B for He ions in 0.16-0.23 MeV/n energy range *vs.* heliographic radial difference between the s/c. The symbols refer to the different periods identified in the text. The quantities $r$ and $p$ denote the linear correlation coefficient and its statistical significance. Solid lines are linear least-squares fit to the data for all time intervals.

**Figure 8.** Panel (A): Ratio between SIT-A and SIT-B 1-hr CIR peak intensities *vs.* ratios between SIT-A and ULEIS (blue symbols), and SIT-B and ULEIS (red symbols) 1-hr CIR peak intensities for He ions in 0.16-0.23 MeV/n energy range. Panel (B): Latitude difference between STEREO-A and STEREO-B divided by HCS tilt angle *vs.* latitude differences between STEREO-A and ACE (blue symbols) and STEREO-B and ACE (red symbols).

**Table 1.** CIR events

| | STEREO-A | | | | STEREO-B | | | | | | |
|---|---|---|---|---|---|---|---|---|---|---|---|
| CIR # | Start (day of year) | Peak Intensity (1/cm2 s sr MeV/n) | Peak (day of year) | Peak Latitude (°) | Start (day of year) | Peak Intensity (1/cm2 s sr MeV/n) | Peak (day of year) | Peak Latitude (°) | Carrington rotation | Corrot. Delay (days) | Notes |
| 2007 | | | | | | | | | | | |
| 1 | 29.0 | 365±5 | 30.1 | -6.0 | 29.0 | 703±17 | 29.9 | -5.7 | 2052 | 0.05 | m |
| 2 | 44.0 | 47.2±1.8 | 45.7 | -6.9 | 44.0 | 62.7±2.5 | 45.8 | -6.7 | 2053 | 0.06 | m |
| 3 | 58.5 | 81.8±2.2 | 59.6 | -7.3 | 58.5 | 85.4±2.6 | 59.5 | -7.2 | 2053 | 0.09 | m |
| 4 | 72.0 | 82.6±2.3 | 72.7 | -7.3 | 72.0 | 80.4 ±2.0 | 72.8 | -7.3 | 2054 | 0.1 | m |
| 5 | 91.0 | 16.6±1.0 | 92.4 | -6.4 | 91.0 | 37.9±1.5 | 92.3 | -6.7 | 2054 | 0.2 | m |
| 6 | 98.5 | 16.6±1.0 | 99.4 | -5.9 | 98.5 | 31.0±1.4 | 99.6 | -6.3 | 2055 | 0.3 | m |
| 7 | 127.5 | 192±4 | 128.3 | -2.9 | 127.5 | 379±5 | 128.4 | -4.0 | 2056 | 0.5 | m,a |
| 8 | 138.0 | 21.7±1.2 | 139.1 | -1.5 | 138.0 | 39.2±1.6 | 139.5 | -2.8 | 2056 | 0.6 | m |
| 9 | 144.0 | 150±3 | 144.6 | -0.8 | 144.0 | 337±5 | 144.6 | -2.3 | 2056 | 0.7 | m,ca,cb |
| 10 | 154.0 | 23.5±1.2 | 155.6 | 0.7 | 154.0 | 102±3 | 154.8 | -1.2 | 2057 | 0.8 | m |
| 11 | 192.0 | 337±5 | 192.9 | 5.2 | 191.5 | 1079±15 | 192.4 | 2.9 | 2058 | 1.4 | m,a,b |
| 12 | 202.0 | 1.15±0.39 | 203.5 | 6.1 | 201.0 | 15.1±1.0 | 201.4 | 3.8 | 2058 | 1.5 | m |
| 13 | 219.0 | 7.07±0.69 | 220.2 | 7.1 | 218.0 | 33.2±1.4 | 219.0 | 5.3 | 2059 | 1.8 | m |
| 14 | 239.0 | 38.6±1.6 | 239.4 | 7.3 | 237.5 | 7.09±0.67 | 238.1 | 6.5 | 2060 | 2.1 | m,ca |
| 15 | 264.0 | 54.6±1.9 | 266.2 | 6.2 | 262.0 | 40.6±1.6 | 264.6 | 7.3 | 2061 | 2.4 | m,a |
| 16 | 271.5 | 22.5±1.2 | 272.8 | 5.7 | 269.5 | 45.0±1.7 | 270.9 | 7.3 | 2061 | 2.5 | m,a,b |
| 17 | 298.5 | 17.7±1.1 | 299.4 | 2.9 | 297.5 | 15.6±1.0 | 297.7 | 6.6 | 2062 | 2.8 | m,cb |
| 18 | 318.0 | 12.6±0.9 | 319.4 | 0.3 | 316.5 | 5.76±0.60 | 317.4 | 5.4 | 2063 | 3.0 | m |
| 19 | 325.0 | 84.5±2.3 | 326.3 | -0.6 | 324.0 | 69.3±2.1 | 324.7 | 4.7 | 2063 | 3.1 | m,ca,cb |
| 20 | 346.0 | 41.4±1.6 | 347.1 | -3.3 | 343.0 | 65.2±2.0 | 344.0 | 2.8 | 2064 | 3.2 | m |
| 21 | 353.0 | 35.4±1.5 | 354.9 | -4.1 | 350.0 | 19.1±1.1 | 351.1 | 1.9 | 2064 | 3.3 | m |
| 2008 | | | | | | | | | | | |
| 22 | 6.0 | 50.8±1.8 | 7.9 | -5.9 | 3.5 | 83.3±2.3 | 4.4 | -0.3 | 2065 | 3.3 | m,a,cb |
| 23 | 33.0 | 25.7±1.3 | 35.2 | -7.3 | 29.5 | 20.7±1.1 | 30.6 | -3.5 | 2066 | 3.4 | m |
| 24 | 42.5 | 79.2±2.3 | 43.4 | -7.3 | 40.0 | 3195±56 | 40.8 | -4.6 | 2066 | 3.4 | m,b,cb |
| 25 | 61.0 | 39.9±1.6 | 62.1 | -6.9 | 57.5 | 50.3±1.8 | 58.6 | -6.1 | 2067 | 3.5 | m |
| 26 | 69.5 | 1098±12 | 69.9 | -6.5 | 67.0 | 7.35±0.68 | 67.6 | -6.6 | 2067 | 3.5 | m,a,b,cb |
| 27 | 88.0 | 8.29±0.74 | 89.2 | -5.0 | 85.0 | 25.0±1.3 | 86.6 | -7.3 | 2068 | 3.5 | m,a |
| 28 | 98.0 | 7.84±0.73 | 99.3 | -3.9 | 94.0 | 124±3 | 95.3 | -7.3 | 2068 | 3.6 | m,a,b |
| 29 | 115.5 | 10.4±0.8 | 116.1 | -1.8 | 112.0 | 1.03±0.27 | 113.1 | -6.9 | 2069 | 3.7 | m |
| 30 | 168.5 | 11.4±0.9 | 169.4 | 4.9 | 165.5 | 2.03±0.36 | 166.4 | -2.5 | 2071 | 4.2 | m |
| 31 | 224.0 | 26.5±1.3 | 224.9 | 7.2 | 220.0 | 36.3±1.5 | 220.8 | 3.3 | 2073 | 5.0 | m |
| 32 | 234.0 | 3.23±0.50 | 236.2 | 6.8 | 229.0 | 78.3±2.2 | 230.1 | 4.2 | 2073 | 5.2 | m,cb |
| 33 | 250.5 | 5.58±0.62 | 252.1 | 5.7 | 245.5 | 10.1±0.8 | 246.3 | 5.5 | 2074 | 5.4 | m |
| 34 | 277.5 | 10.3±0.8 | 278.2 | 3.0 | 272.0 | 70.0±2.1 | 273.5 | 7.0 | 2075 | 5.8 | m,b |
| 35 | 305.5 | 111±3 | 306.3 | -0.7 | 299.0 | 2.34±0.39 | 302.1 | 7.2 | 2076 | 6.1 | m,ca |
| 36 | 314.0 | 35.0±1.5 | 314.9 | -1.8 | 309.5 | 1.27±0.29 | 309.9 | 7.1 | 2076 | 6.2 | |
| 37 | 333.5 | 120±3 | 335.0 | -4.2 | 327.0 | 22.3±1.2 | 328.4 | 6.3 | 2077 | 6.4 | ca |
| 38 | 350.5 | 1.49±0.47 | 352.4 | -5.9 | 344.0 | 53.3±1.8 | 346.5 | 4.9 | 2077 | 6.5 | |
| 2009 | | | | | | | | | | | |
| 39 | 48.5 | 53.4±1.9 | 48.9 | -6.5 | 41.5 | 1.83±0.36 | 46.9 | -2.7 | 2079 | 6.8 | a |
| 40 | 74.5 | 0.66±0.41 | 75.6 | -4.2 | 69.0 | 74.7±2.2 | 69.5 | -5.1 | 2080 | 6.9 | b,x |
| 41 | 83.0 | 0.51±0.43 | 83.5 | -3.2 | 76.5 | 10.2±0.8 | 76.9 | -5.7 | 2081 | 6.9 | x |
| 42 | 129.5 | 0.57±0.34 | 132.8 | 3.2 | 122.0 | 22.2±1.2 | 123.0 | -7.2 | 2082 | 7.2 | x |
| 43 | 145.0 | 0.66±0.38 | 147.7 | 4.8 | 140.0 | 23.9±1.2 | 141.4 | -6.6 | 2083 | 7.3 | x |
| 44 | 152.5 | 139±3 | 153.6 | 5.4 | 146.5 | 0.52±0.26 | 146.9 | -6.3 | 2083 | 7.3 | ca,x |
| 45 | 175.0 | 12.9±0.9 | 176.0 | 7.0 | 170.0 | 0.64±0.26 | 171.1 | -4.4 | 2084 | 7.5 | cb,x |
| 46 | 207.5 | 3.32±0.48 | 207.8 | 7.1 | 199.5 | 300±5 | 200.0 | -1.4 | 2085 | 7.9 | b,cb |
| 47 | 260.5 | 0.41±0.34 | 260.8 | 2.5 | 252.5 | 20.3±1.1 | 253.5 | 4.2 | 2087 | 8.7 | cb,x |

Notes: (1) Each row indicates two energetic ion events observed on STEREO-A and STEREO-B associated with single CIR. (2) The peak intensities (1-hr averaged) are from the energy range 0.16-0.23 MeV/n.

(a) -reverse shock at CIR observed by STEREO-A
(b) -reverse shock at CIR observed by STEREO-B
(ca) -mixed ICME-CIR event on STEREO-A
(cb) -mixed ICME-CIR event on STEREO-B
(m) -event in survey of *Mason et al.* [2009]
(x) -very small event either on STEREO-A or STEREO-B

**Table 2.** ACE CIR events

| CIR # | Start (day of year) | Peak Intensity (1/cm2 s sr MeV/n) | Peak (day of year) | Peak Latitude (°) |
|---|---|---|---|---|
| 2007 | | | | |
| 1 | 29.5 | 283±13 | 30.1 | -5.9 |
| 2 | 44.0 | 53.6±3.4 | 45.2 | -6.8 |
| 3 | 58.5 | 87.4±4.5 | 59.6 | -7.2 |
| 4 | 72.0 | 114±5 | 72.7 | -7.2 |
| 5 | 91.0 | 21.4±2.1 | 92.1 | -6.5 |
| 6 | 98.5 | 14.6±1.7 | 99.5 | -6.1 |
| 7 | 127.5 | 296±13 | 128.2 | -3.5 |
| 8 | 138.0 | 27.0±2.5 | 138.8 | -2.3 |
| 9 | 144.0 | 277±13 | 144.6 | -1.6 |
| 10 | 154.0 | 78.1±4.1 | 154.9 | -0.4 |
| 11 | 192.0 | 362±18 | 192.4 | 3.9 |
| 12 | 202.0 | 4.12±0.92 | 202.5 | 4.9 |
| 13 | 219.0 | 13.5±1.7 | 219.5 | 6.2 |
| 14 | 238.0 | 5.36±1.05 | 238.9 | 7.1 |
| 15 | 264.0 | 57.3±3.5 | 265.1 | 7.1 |
| 16 | 270.5 | 10.9±1.5 | 272.6 | 6.8 |
| 17 | 298.0 | 94.4±4.8 | 298.8 | 5.0 |
| 18 | 317.0 | 15.6±1.8 | 317.8 | 3.0 |
| 19 | 324.0 | 487±23 | 325.4 | 2.1 |
| 20 | 344.0 | 49.8±3.2 | 345.9 | -0.5 |
| 21 | 351.0 | 433±18 | 352.4 | -1.3 |
| 2008 | | | | |
| 22 | 4.5 | 42.1±3.0 | 6.1 | -3.5 |
| 23 | 31.5 | 15.7±1.8 | 32.8 | -6.0 |
| 24 | 41.0 | 344±15 | 41.9 | -6.6 |
| 25 | 58.0 | 26.1±2.3 | 60.6 | -7.2 |
| 26 | 69.0 | 38.2±2.8 | 69.3 | -7.2 |
| 27 | 86.0 | 20.6± 2.1 | 87.6 | -6.8 |
| 28 | 96.0 | 12.2±1.6 | 97.3 | -6.2 |
| 29 | 113.5 | 10.5±1.5 | 114.5 | -4.9 |
| 30 | 167.0 | 2.68±0.74 | 168.3 | 1.2 |
| 31 | 222.0 | 573±30 | 222.9 | 6.4 |
| 32 | 231.0 | 4.48±0.97 | 231.4 | 6.8 |
| 33 | 247.5 | 2.70±0.75 | 248.4 | 7.2 |
| 34 | 275.0 | 40.3±2.9 | 275.7 | 6.7 |
| 35 | 302.5 | 12.7±1.9 | 303.9 | 4.6 |
| 36 | 312.5 | 5.43±1.07 | 313.2 | 3.6 |
| 37 | 330.0 | 308±13 | 330.9 | 1.5 |
| 38 | 346.5 | 1.65±0.58 | 348.4 | -0.7 |
| 2009 | | | | |
| 39 | 45.0 | 21.0±2.5 | 45.6 | -6.8 |
| 40 | 72.0 | 5.38±1.27 | 72.5 | -7.2 |
| 41 | 79.0 | 0.89±0.51 | 80.7 | -7.0 |
| 42 | 125.5 | 2.17± 0.82 | 126.6 | -3.6 |
| 43 | 143.0 | 8.46±1.63 | 145.3 | -1.5 |
| 44 | 149.0 | 2.51± 0.89 | 150.2 | -0.9 |
| 45 | 171.0 | 1.28± 0.75 | 172.6 | 1.8 |
| 46 | 203.0 | 35.1± 3.4 | 204.4 | 5.1 |
| 47 | 257.0 | 0.34± 0.34 | 258.4 | 7.2 |

Note: The peak intensities (1-hr averaged) are from the energy range 0.16-0.23 MeV/n.

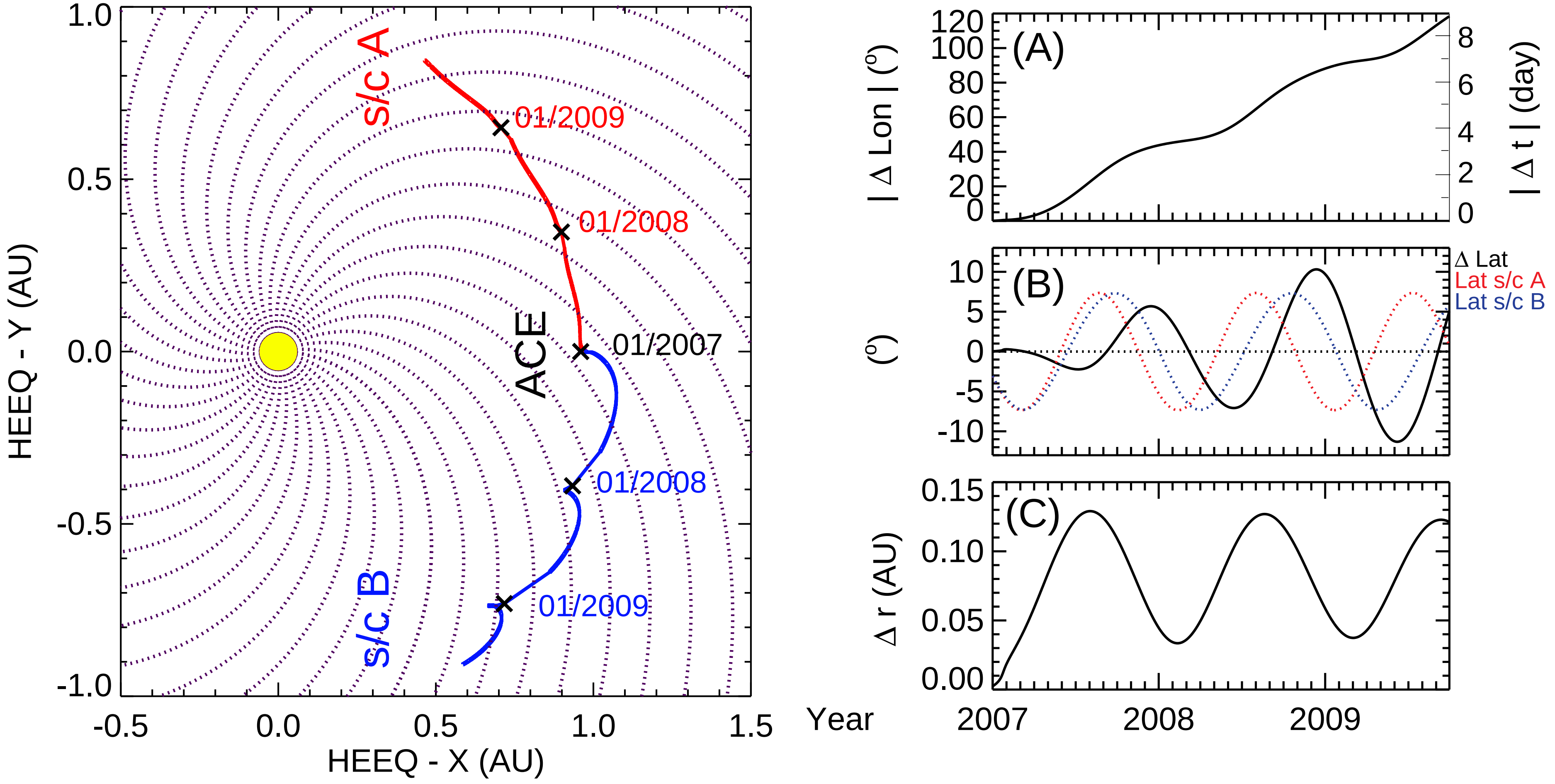

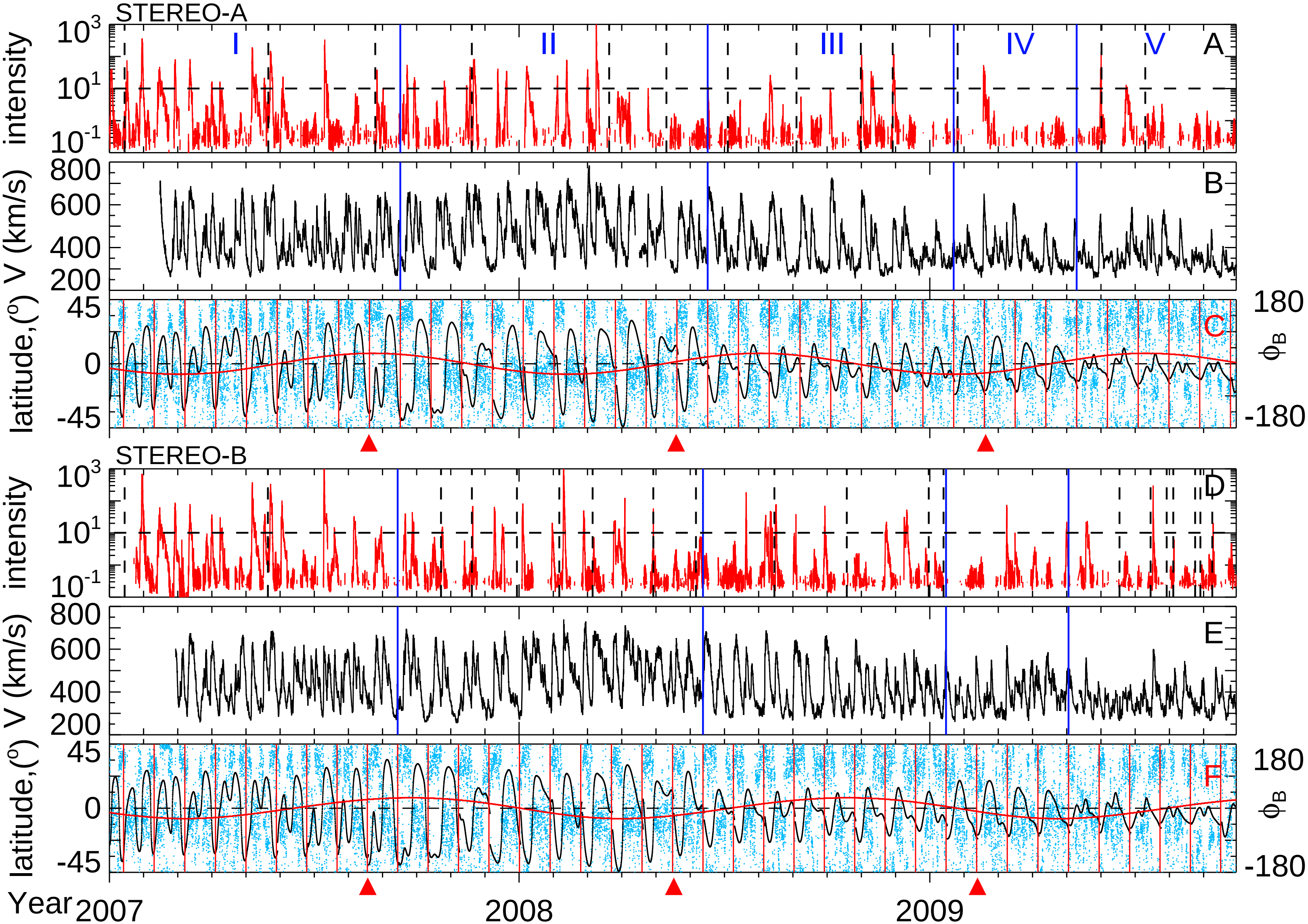

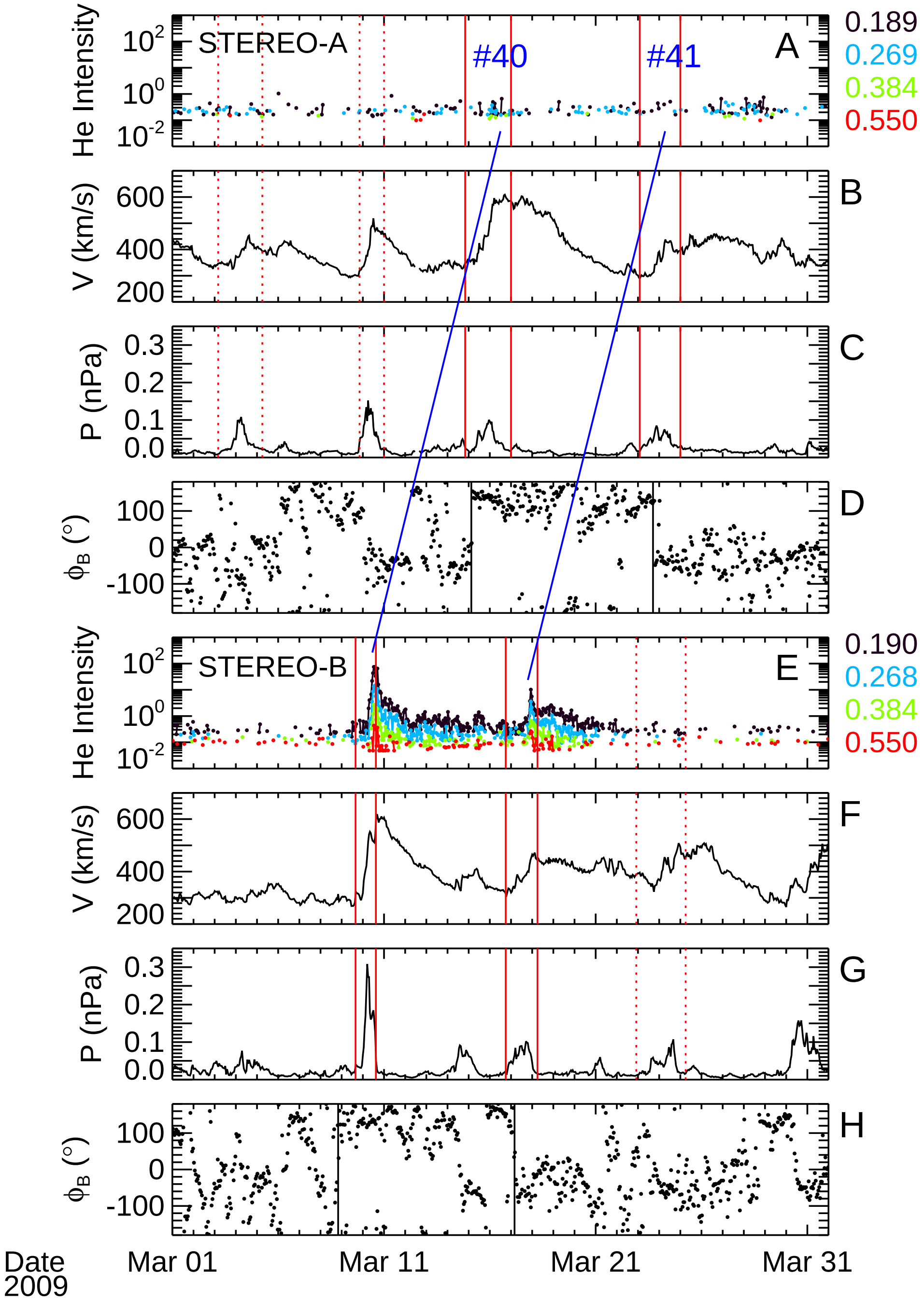

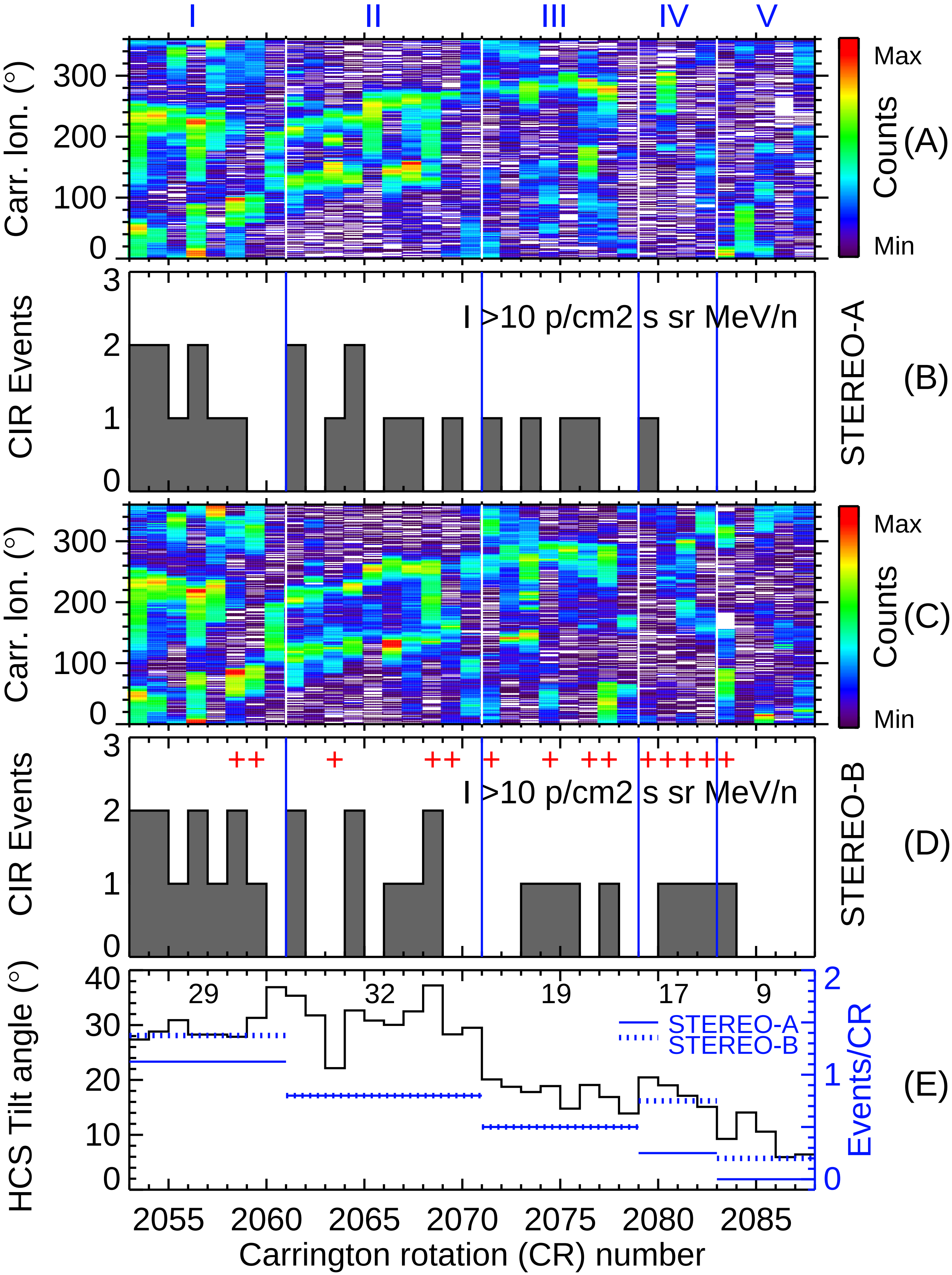

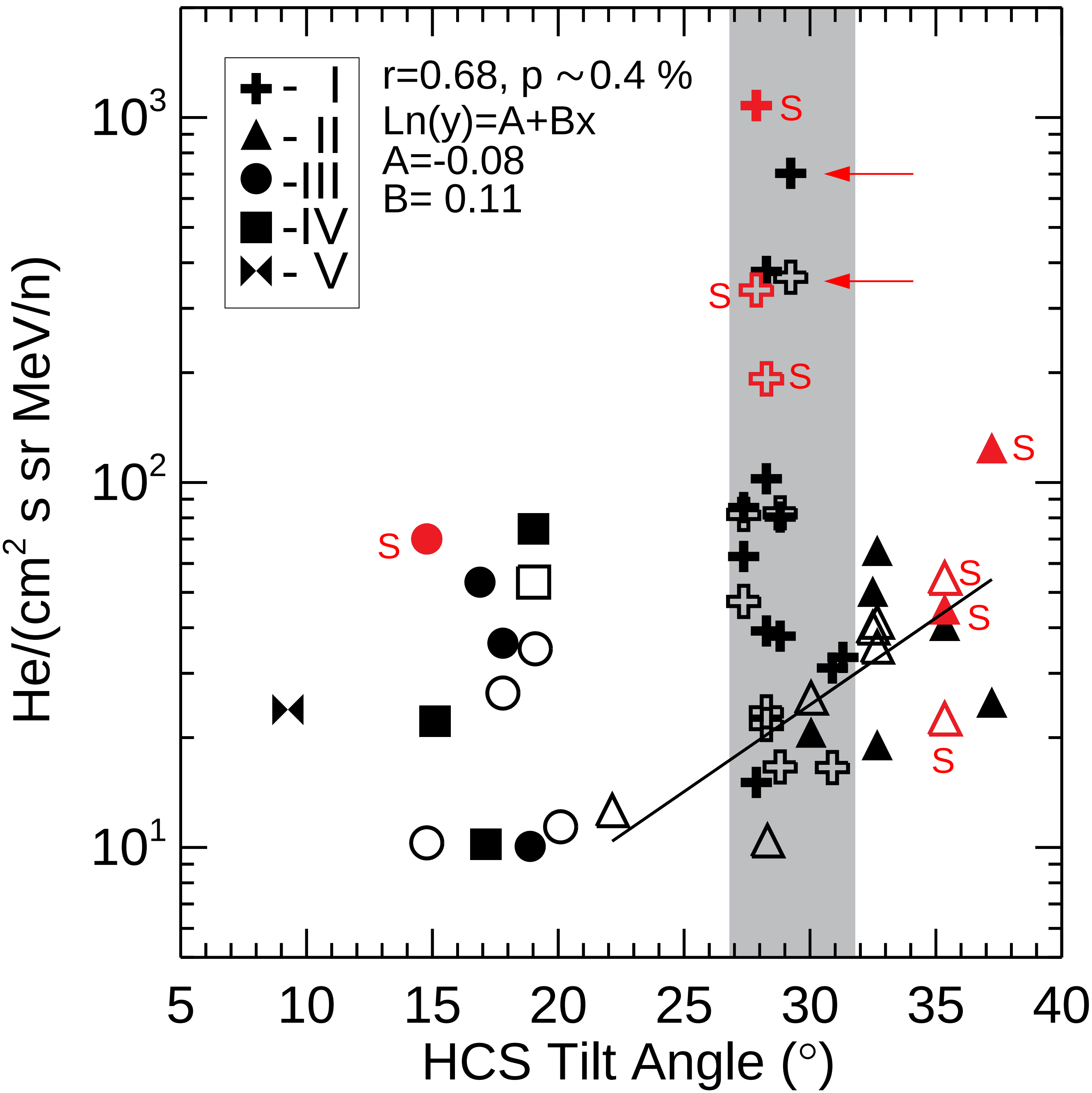

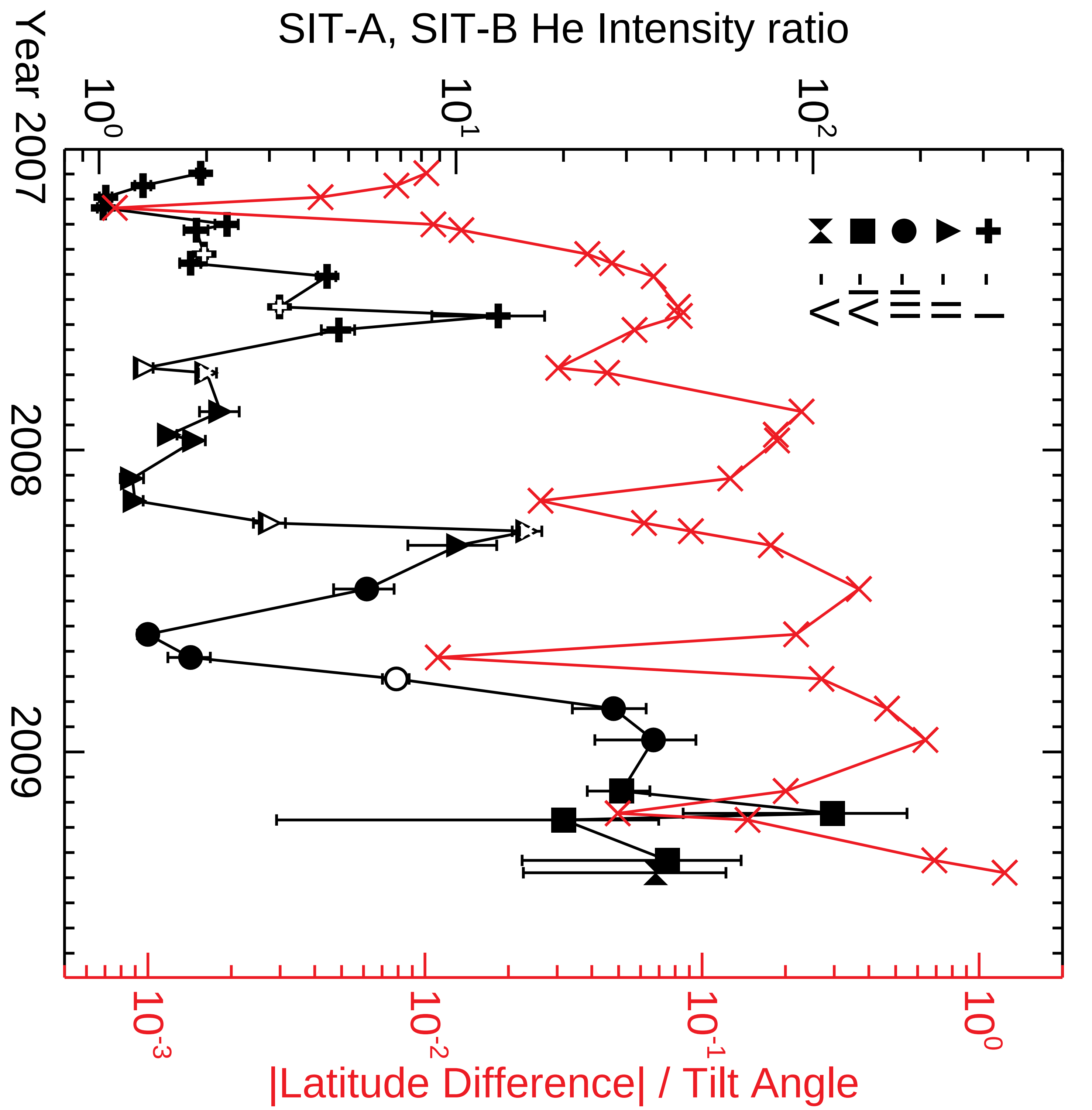

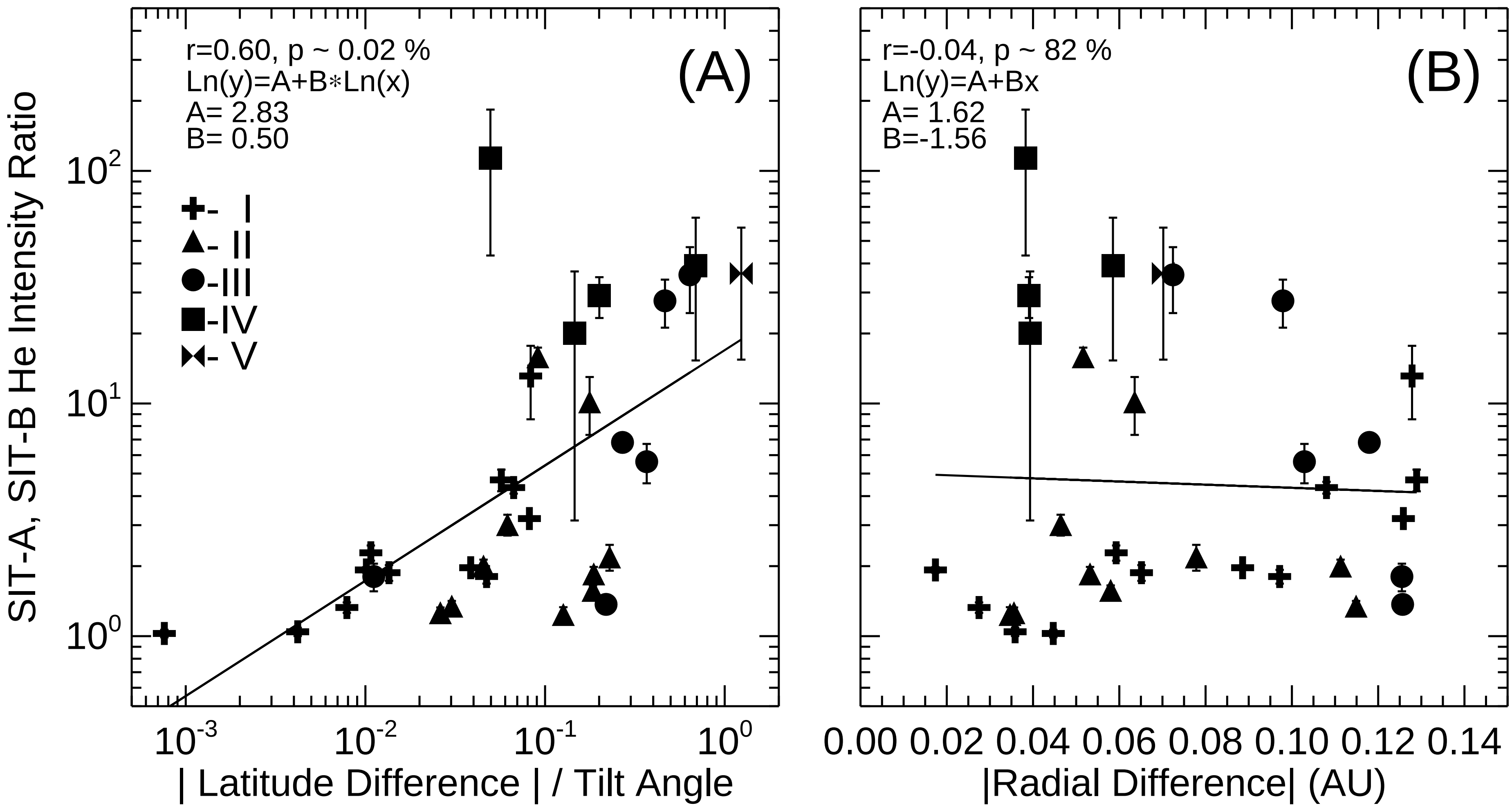

r=0.60, p ~ 0.02 %
Ln(y)=A+B∗Ln(x)
A= 2.83
B= 0.50
(A)
+- I
▲- II
●-III
■-IV
⋈- V
SIT-A, SIT-B He Intensity Ratio
$10^2$
$10^1$
$10^0$
$10^{-3}$
$10^{-2}$
$10^{-1}$
$10^0$
| Latitude Difference | / Tilt Angle
r=-0.04, p ~ 82 %
Ln(y)=A+Bx
A= 1.62
B=-1.56
(B)
0.00 0.02 0.04 0.06 0.08 0.10 0.12 0.14
|Radial Difference| (AU)

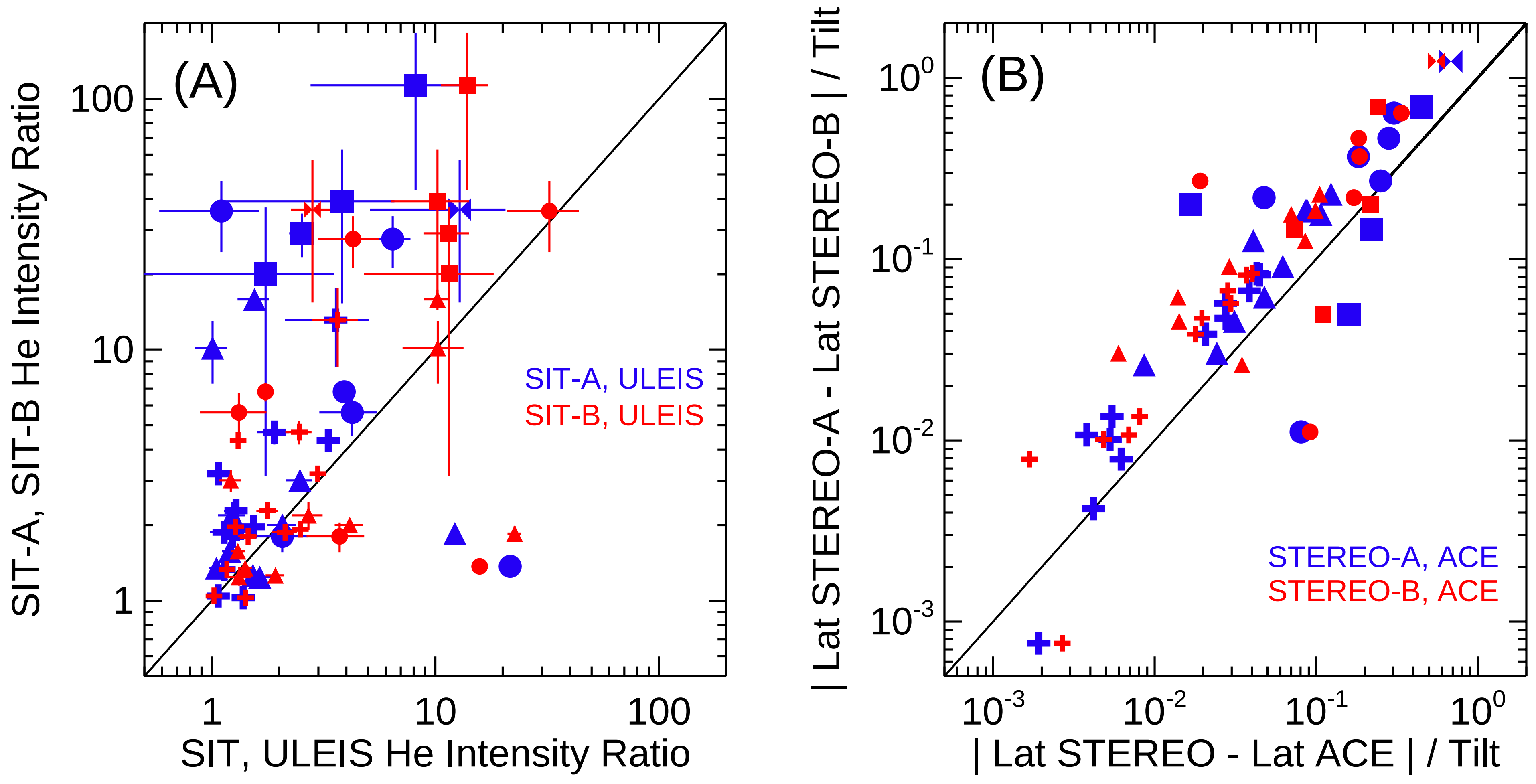
(A)
100
10
1
1
10
100
SIT-A, SIT-B He Intensity Ratio
SIT, ULEIS He Intensity Ratio
SIT-A, ULEIS
SIT-B, ULEIS
(B)
$10^0$
$10^{-1}$
$10^{-2}$
$10^{-3}$
| Lat STEREO-A - Lat STEREO-B | / Tilt
| Lat STEREO - Lat ACE | / Tilt
STEREO-A, ACE
STEREO-B, ACE